\documentclass[hyper,a4paper]{SIMO}  
\usepackage{graphicx}
\usepackage{epsfig,rotating,multirow}
\usepackage{exscale}
\usepackage{amsmath}
\usepackage{cite}
\usepackage{latexsym}
\usepackage{graphics}
\usepackage[english]{babel}

\newcommand{\ve}{\varepsilon}

\newcommand{\bm}{\begin{minipage}}
\renewcommand{\em}{\end{minipage}}
\newcommand{\bi}{\begin{itemize}}
\newcommand{\ei}{\end{itemize}}
\newcommand{\be}{\begin{eqnarray}}
\newcommand{\en}{\end{eqnarray}}
\newcommand{\ba}{\begin{array}}
\newcommand{\ea}{\end{array}}
\newcommand{\bc}{\begin{center}}
\newcommand{\ec}{\end{center}}
\newcommand{\bfi}{\begin{figure}}
\newcommand{\efi}{\end{figure}}

\newcommand{\re}{\mbox{Re}}
\newcommand{\bfr}{\begin{flushright}}
\newcommand{\efr}{\end{flushright}}

\newcommand{\no}{\nonumber}
\newcommand{\ra}{\rightarrow}

\newcommand{\ov}[1]{\overline{#1}}

\renewcommand{\=}[1]{\!\!\!&{#1}&\!\!\!}

\newcommand{\bbr}{{\it B\scalebox{.8}{A}B\scalebox{.8}{AR}} }

\newcommand{\phipiz}{\phi\pi^0}
\title{
Study of the $\phipiz$ transition form factor
}
\author{Simone Pacetti\\
Enrico Fermi Center, Rome, Italy\\
INFN, Laboratori Nazionali di Frascati, Frascati, Italy  \\ 
E-mail: 
\email{simone.pacetti@lnf.infn.it} }
\abstract{%
Recently the \bbr\ Collaboration published new data on the cross section 
for the annihilation $e^+e^-\to\phipiz$, obtained using the initial 
state radiation technique at a center of mass energy of 10.6 GeV. 
\\
Such a process represents an interesting test bed for the quark model.
Indeed, since the $\phipiz$ production via $e^+e^-$ annihilation proceeds 
through a mechanism which violates the Okubo-Zweig-Iizuka rule, the 
corresponding cross section could be characterized by contributions 
from non-$q\ov{q}$ bound states, like hybrids or tetraquarks.
\\
The $\phipiz$ cross section is analyzed in connection with other data coming
from different processes, that involve the same mesons, using a method
which implements the analyticity in the $\phipiz$ transition form factor 
by means of a dispersion relation procedure.%
}
\begin{document}
%
%
%%%%%%%%%%%%%%%%%%%%%%%%%%%%%%%%%%%%%%%%%%%%%%%%%%%%%%%%%%%%%%%%%%%%%%%
%
% 
%
\section{Introduction}
\label{sec:intro}
Recently the \bbr\ Collaboration measured for the first time
the cross section for the annihilation $e^+e^-\to\phipiz$~\cite{babar}.
The data have been achieved using the initial state radiation technique
that allows to scan energies, for the invariant mass of the hadronic final 
state, from the production threshold up to about 4.6 GeV.
\\
There have been some previous attempts~\cite{russi} to observe such a final 
state in $e^+e^-$ annihilation but only upper limits were given.
\\
The process $e^+e^-\to\phipiz$ plays a crucial role in the understanding
of the quark model. Indeed, the $\phipiz$ channel, which is forbidden by the 
Okubo-Zweig-Iizuka (OZI)~\cite{ozi} rule for any $q\ov{q}$ vector meson, 
could be a likely decay mode for non-standard bound states as
hybrids and tetraquarks~\cite{Close:1978be}.
\\
The $\phipiz$ cross section is analyzed in connection with other data coming
from different processes, that involve the same mesons, using a method
which implements the analyticity in the $\phipiz$ transition form factor (TFF) 
by means of a dispersion relation procedure.%
\\
This analysis is based on the method reported in Ref.~\cite{mia}. 
Such a method, which will be described in the following, 
has the advantage of using TFF's instead
of cross sections and decay rates. A TFF is defined in different energy
regions and is experimentally accessible through different processes.
In particular the $\phipiz$ TFF can be parametrized in terms of resonant
contributions, whose coupling with the mesons under consideration is inferred
from other measurements. In this view, the OZI rule violation can be seen as a
direct consequence of the large $\phi-\rho^0\pi^0$ coupling
(BR$(\phi\to3\pi)\simeq 15\%$~\cite{pdg}).\\
In addition, since we use dispersion relations (DR) to implement analyticity,
we need to know the asymptotic behavior of the TFF, for which we adopt 
the quark-counting rule (QcR) prescription~\cite{brodsky}. As we will see in 
Sec.~\ref{sec:tff-def}, this provides a
further constraint on the quark structure of the involved mesons.
\\
The paper is organized as follows. In Sec.~\ref{sec:TFF} we introduce the main
formulae for the cross section and decay rates, and a general analytic description
for the $\phipiz$ TFF, whose free parameters are the properties of the 
vector meson contributions in the low-energy region. 
In Sec.~\ref{sec:rho0} we discuss a first case where for the description of the
$\phipiz$ TFF we consider only one 
contribution, the $\rho^0$. Since such an intermediate state is completely 
known through the decay $\phi\to\rho^0\pi^0$, there are no degrees of freedom,
and the description, in this case, is model-independent. In Sec.~\ref{sec:global} we develop this
procedure including further intermediate states, essentially $\rho$-recurrences, to
achieve a global description of the TFF. Finally, the obtained results are 
summarized and discussed in the last section. 
\section{$\phipiz$ transition form factor}
\label{sec:TFF}
\subsection{Cross section and decay rate formulae}
\label{sec:formulae}
To extract the TFF values, data on cross section and decay rates
have to be compared with the expressions for these quantities
obtained under the assumption of pointlike mesons. 
\bfi[ht!]
\bc
\epsfig{file=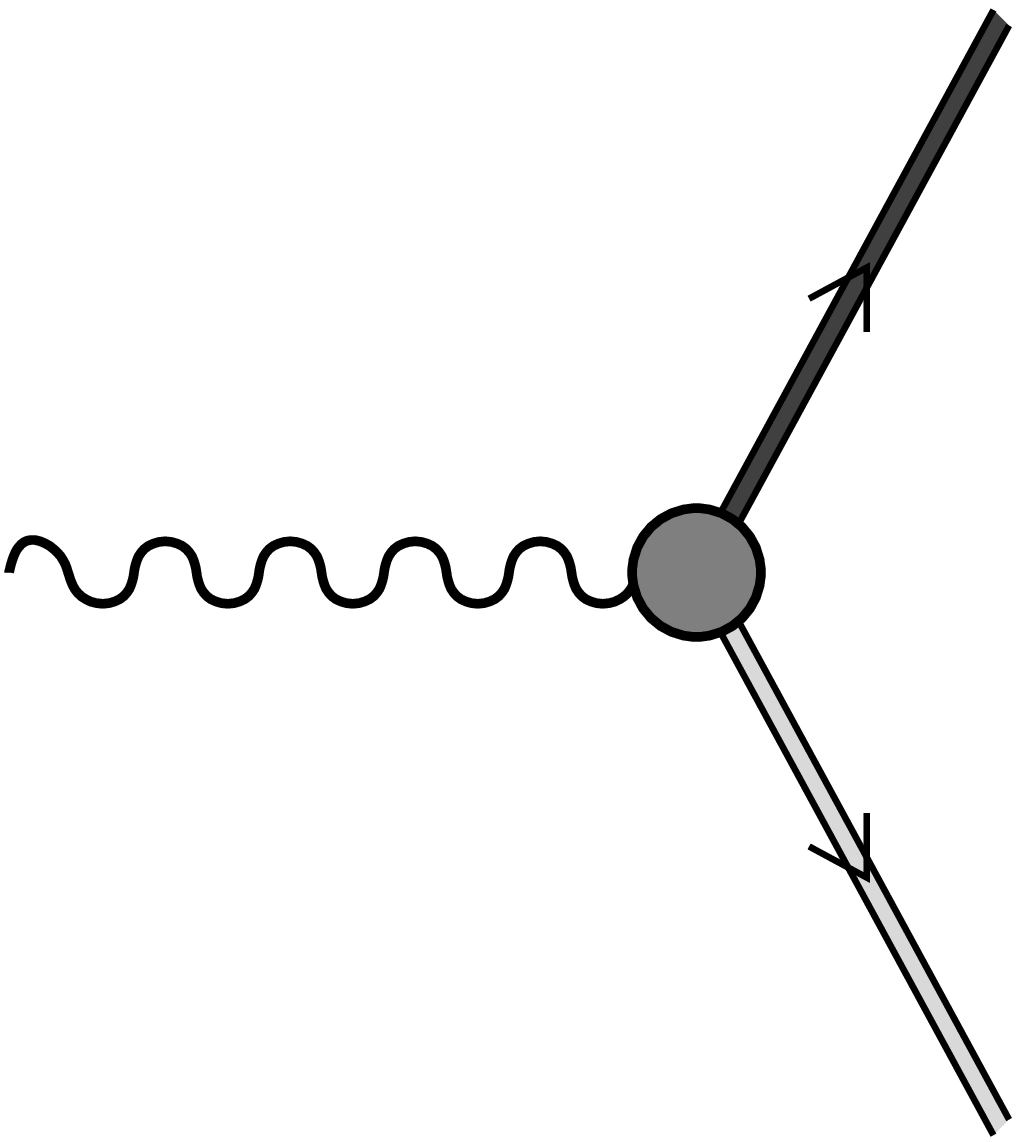,height=30mm}
\put(-101,42){$\gamma^*(q)$}
\put(-0,-2){$\pi^0(k)$}
\put(-0,80){$\phi(p,\ve)$}
\put(-15,38){$F_{\phipiz}(q^2)$}
\caption{\label{fig:fey-phipi0}
Diagram of the conversion $\phi\gamma\pi^0$.
}
\ec
\efi\vspace{-0mm}\\
The conversion
current for the vertex of fig.~\ref{fig:fey-phipi0}, being $\phi$  
a vector and $\pi^0$ a pseudoscalar meson, has the general form:
\be 
J_{\phipiz}^\mu=e\,\varepsilon^{\mu\nu\rho\sigma}\epsilon_\nu p_\rho q_\sigma
\,F_{\phipiz}(q^2)\;,
\label{eq:current}
\en
where $e$ is the electron charge, 
$\varepsilon^{\mu\nu\rho\sigma}$ is the fully antisymmetric Levi-Civita 
tensor, $q$ is the 4-momentum of the photon and, $p$ and $\epsilon$ are the 
4-momentum and the polarization vector of the $\phi$. The tensor structure of 
this current follows from Lorentz and gauge invariance.\\
The TFF $F_{\phipiz}(q^2)$ is an analytic function defined in the $q^2$ complex 
plane with the cut $(4M_\pi^2,\infty)$ along the real axis. Only real values 
of $q^2$ are experimentally accessible. The TFF is real for $q^2$ below the 
threshold $4M_\pi^2$, while is complex over the cut, i.e. for $q^2>4M_\pi^2$. 
It describes the
photon-hadron vertex in terms of electromagnetic interaction with the quark
constituents. The underlying physical process is the $\phi$ radiative decay:
$\phi\to \pi^0\gamma$, which occurs at $q^2=0$. 
There are two other one-photon exchange processes, which involve the
same conversion, that can be used to investigate the $\phipiz$ TFF 
(see fig.~\ref{fig:q2-axis}):
\bi
\item the decay: $\phi\to \pi^0 e^+e^-$ with $4m_e^2<q^2<(M_\phi-M_{\pi^0})^2$;
\item the annihilation: $e^+e^-\to\phipiz$ with $q^2\ge(M_\phi+M_{\pi^0})^2 $.
\ei
\bfi[h!]
\bc
\epsfig{file=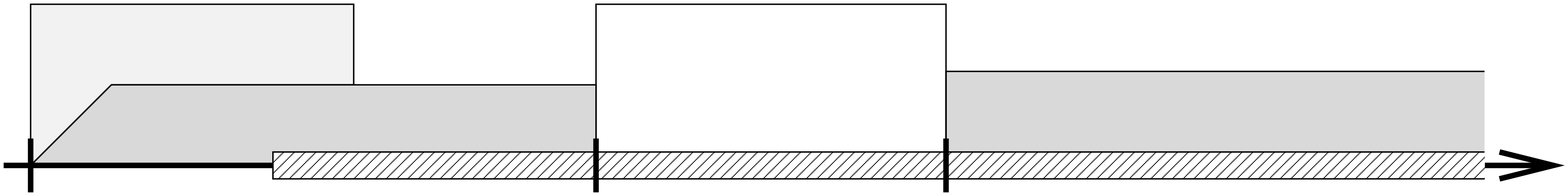,width=120mm}
\put(-325,30){\boldmath$\phi\to\pi^0\gamma$}
\put(-300,13){\boldmath$\phi\to\pi^0 e^+e^-$}
\put(-200,29){unphysical}
\put(-190,16){region}
\put(-110,14){\boldmath$e^+e^-\to\phipiz$}
\put(-7,-6){\boldmath$\re\, q^2$}
\put(-237,-8){$(M_\phi\!-\!M_{\pi^0})^2$}
\put(-160.5,-8){$(M_\phi\!+\!M_{\pi^0})^2$}
\put(-290,-8){$4M_{\pi^0}^2$}
\put(-337,-8){0}
\caption{\label{fig:q2-axis}
The $q^2$ real axis (not in scale). The three processes, involving the mesons $\phi$ and $\pi^0$,
in the corresponding regions are also indicated. The lined band represents the
cut, where the TFF is complex.}
\ec
\efi\vspace{-0mm}
Rates and cross section can be computed, as functions of the TFF, using
the current of eq.~(\ref{eq:current}). The radiative decay rate is the constant
quantity:
\be
\Gamma=
\frac{\alpha}{3}\left(\frac{M_\phi^2-M_{\pi^0}^2}{2M_\phi}\right)^3
[F_{\phipiz}(0)]^2\;,
\label{eq:rate-rad}
\en
it is proportional to the squared value of the TFF at $q^2=0$.\\
The differential rate for the conversion decay $\phi\to \pi^0 e^+e^-$ 
has the form
\be
\frac{d\Gamma}{dq^2}(q^2)\!=\!
\frac{\alpha^2}{9\pi}
\sqrt{1\!-\!\frac{4m_e^2}{q^2}}
\left(1\!+\!\frac{2m_e^2}{q^2}\right)
\frac{1}{q^2}
\left[\frac{
\big(q^2\!+\!M_\phi^2\!-\!M_{\pi^0}^2\big)^2\!-\!4M_\phi^2 q^2}{(2M_\phi)^2}
\right]^\frac{3}{2}
|F_{\phipiz}(q^2)|^2\,.
\label{eq:rate-conv}
\en
This is an energy-dependent quantity which gives information on the real value
of the TFF up to $q^2=4M_\pi^2$ and on its modulus in the region
$4M_\pi^2<q^2<(M_\phi-M_{\pi^0})^2$.\\ 
Finally, the annihilation cross section:
\be
\sigma(q^2)\!=\!\frac{\pi}{6}\frac{\alpha^2}{(q^2)^3}
\frac{q^2+2m_e^2}{\sqrt{q^2(q^2-4m_e^2)}}
\Big[\big(q^2\!+\!M_\phi^2\!-\!M_{\pi^0}^2\big)^2\!-\!4M_\phi^2 q^2\Big]^\frac{3}{2}
|F_{\phipiz}(q^2)|^2\,,
\label{eq:rate-annihi}
\en
from which we extract the modulus of the TFF above the physical threshold, i.e. for
$q^2~>~(M_\phi+M_{\pi^0})^2$.
\subsection{The data}
\label{sec:data}
There are only two available sets of data on the $\phipiz$ TFF,
in addition to the recent cross section measurement~\cite{babar},
only the radiative decay rate $\Gamma(\phi\to\pi^0\gamma)$ is
known~\cite{pdg}. Unfortunately, there are no data on $\phi\to\pi^0 e^+e^-$.
The measurement of 
such a rate should cover a wide energy range below the physical
threshold. {\bf The unphysical region} (see figs.~\ref{fig:q2-axis} and~\ref{fig:tff-data}b)
{\bf\boldmath for this TFF, that is the energy interval not experimentally accessible, 
is very narrow: $2M_{\pi^0}\simeq 270$ MeV and no structure is expected in this energy range}.  
From the radiative decay rate we extract the value at zero of the TFF as:
\be
F_{\phipiz}(0)\equiv |g^\phi_{\pi^0\gamma}|=\sqrt{\frac{3\Gamma(\phi\to\pi^0\gamma)}{\alpha}}
\left(\frac{2M_\phi}{M_\phi^2-M_{\pi^0}^2}\right)^{3/2}=
(0.131\pm0.004)\,\rm GeV^{-1}\,,
\label{eq:tff0}
\en 
this value corresponds to the modulus of the coupling $g^\phi_{\pi^0\gamma}$.\\
Finally, from the annihilation cross section data (fig.~\ref{fig:tff-data}a), 
using the expression of eq.~(\ref{eq:rate-annihi}), 
we obtain the TFF above the physical threshold, which is shown in 
fig.~\ref{fig:tff-data}b.
\bfi[ht!]
\bc
\epsfig{file=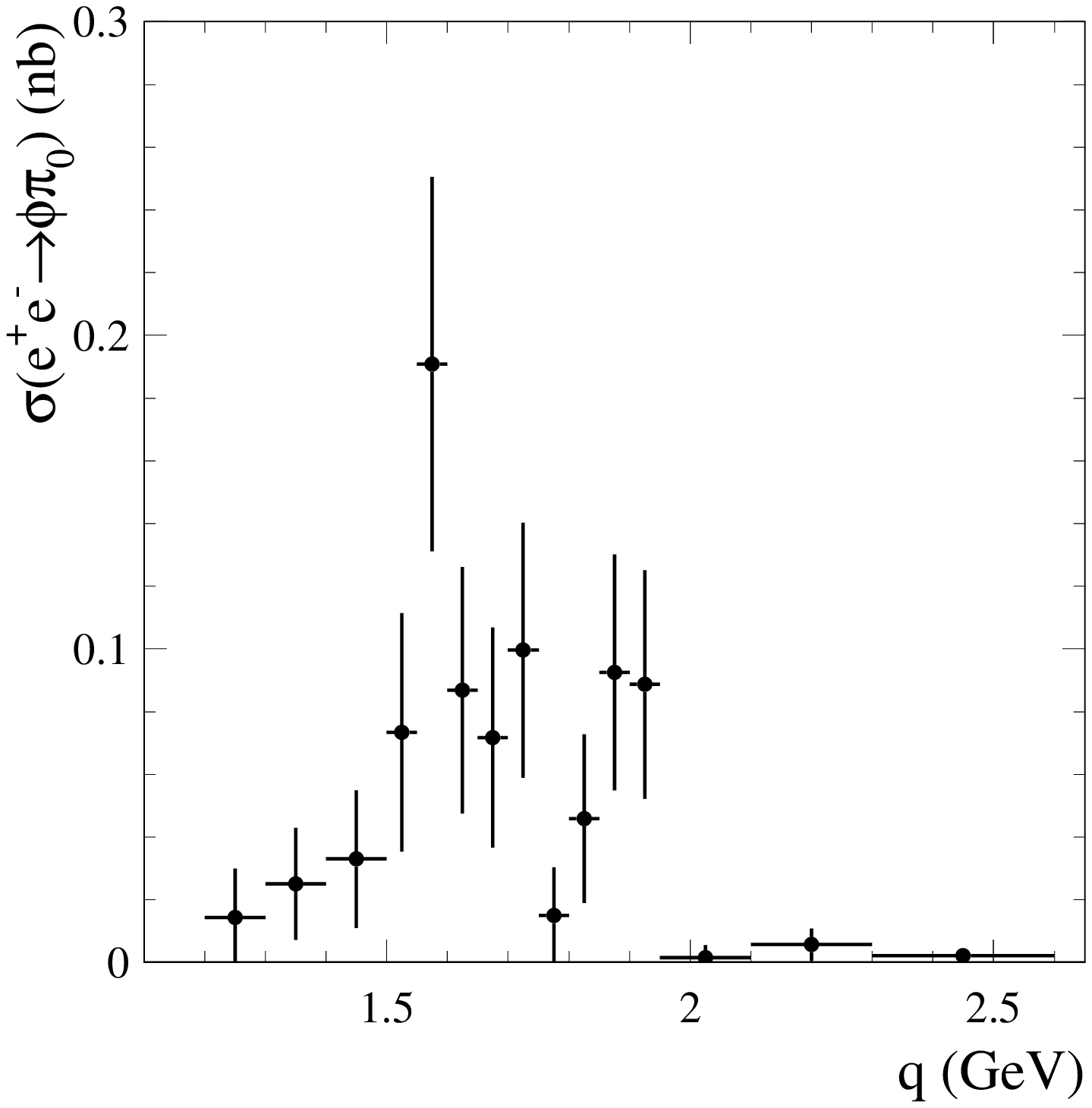,width=70mm}
\put(-165,180){\bf a}
\epsfig{file=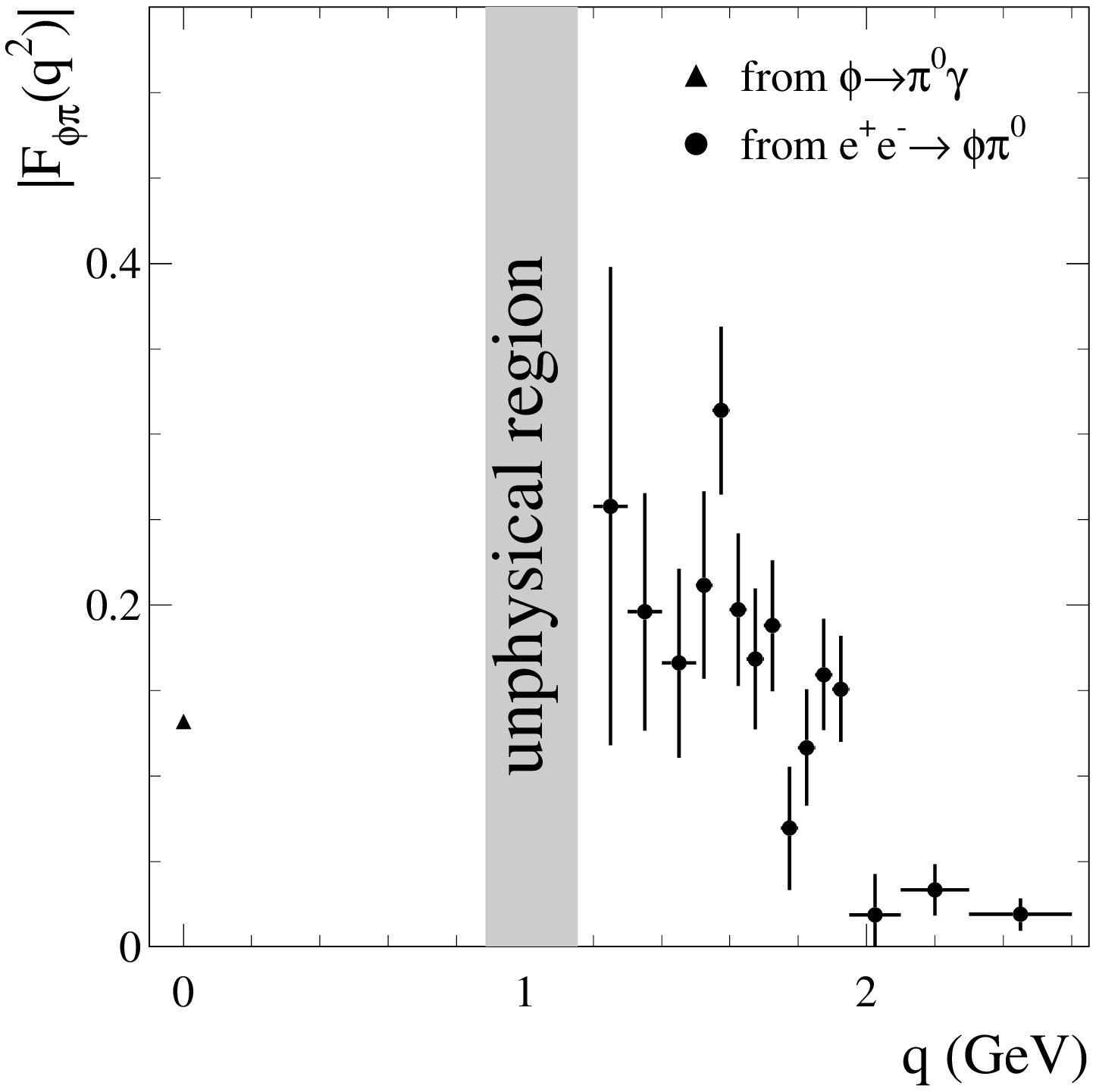,width=70mm}
\put(-165,180){\bf b}
\caption{\label{fig:tff-data}
a) The $e^+e^-\to\phipiz$ cross section~\cite{babar}. b) The $\phipiz$ TFF
extracted from cross section (solid circles) and radiative decay rate (triangle
at $q=0$).}
\ec
\efi\vspace{-0mm}\\
Even though there are no data on the differential rate 
$\frac{d\Gamma(\phi\to\pi^0 e^+e^-)}{dq^2}$, in Ref.~\cite{pdg}
is reported the branching fraction
\be
{\rm BR}(\phi\to\pi^0 e^+e^-)_{\rm PDG}=(1.12\pm0.28)\times 10^{-5}\,,
\en 
this value is the average of the only two existing sets of
data collected by SND and CMD-2 Collaborations~\cite{diff-data1,diff-data2}.
In both cases the cuts applied to extract the events select
small values of $q^2$ and hence do not allow to study the
energy dependence of the TFF as described in eq.~(\ref{eq:rate-conv}).
The branching ${\rm BR}(\phi\to\pi^0 e^+e^-)_{\rm PDG}$ can be interpreted as
the yield of the events with a small $e^+e^-$ invariant 
mass~\cite{diff-data1,diff-data2}. The only information that can be drawn is
about the mean value of the modulus of the TFF  over a small $q^2$ interval,
$[(2m_e)^2,(2m_e+\Delta E_e)^2]$, from the threshold up to
a not claimed maximum energy $\sqrt{q^2}=2m_e+\Delta E_e$.
More in detail this mean value, depending on $\Delta E_e$, can 
be defined as:
\be
\Big|\ov{F}_{\phipiz}^{\Delta E_e}\Big|^2=
\frac{\Gamma(\phi\to\pi^0 e^+e^-)_{\rm PDG}}
{\frac{\alpha^2}{9\pi}{\displaystyle\int}_{(2m_e)^2}^{(2m_e+\Delta E_e)^2}
\!\!\!\!\sqrt{1\!-\!\frac{4m_e^2}{q^2}}
\left(1\!+\!\frac{2m_e^2}{q^2}\right)
\frac{1}{q^2}
\bigg[\frac{
\big(q^2\!+\!M_\phi^2\!-\!M_{\pi^0}^2\big)^2\!\!-\,4M_\phi^2 q^2}{(2M_\phi)^2}
\bigg]^\frac{3}{2}dq^2}~,
\en
for the phase-space integration, at denominator, we use the 
definition given in eq.~(\ref{eq:rate-conv}). With $\Delta E_e=400$ MeV~\cite{diff-data1}
one gets:
\be
\Big|\ov{F}_{\phipiz}^{\Delta E_e=400\;\rm MeV}\Big|=0.14\pm 0.02\;{\rm GeV}^{-1} ~.
\label{eq:average}
\en
This value has to be compared with the TFF at $q^2=0$ of eq.~(\ref{eq:tff0}), 
extracted from the radiative decay rate $\Gamma(\phi\to\pi^0\gamma)$. 
They are perfectly compatible, since the higher value obtained for the
mean value reflects the rising behavior, towards the $\rho^0$ peak, of 
the TFF. However, due to the lack of information on their $q^2$-dependence,
these data will not be used in the following analysis.
\subsection{Parameterization of the transition form factor}
\label{sec:tff-def}
Following the prescriptions given in Ref.~\cite{mia}, the TFF is parameterized
by means of the threefold expression:
\be
F_{\phipiz}(q^2)=\left\{
\begin{array}{ll}
F^{\rm an}_{\phipiz}(q^2) & q^2<4M_\pi^2\\
F^{\rm Res}_{\phipiz}(q^2) & 4M_\pi^2 \le q^2< s^{\rm asy}\\
F^{\rm asy}_{\phipiz}(q^2) & q^2\ge s^{\rm asy}\\
\end{array}\right. ,
\label{eq:3def}
\en
$s^{\rm asy}$, the energy from which we assume the QcR power law behavior, is
a free parameter. 
The three definitions refer to three intervals which cover the whole
time-like region. In more detail:
\bi
\item The resonance region (superscript ``Res''): $4M_\pi^2 \le q^2< s^{\rm asy}$, 
      where the TFF is described in terms of intermediate vector meson resonances ($V_j$
      with $j=1,\ldots,N$).
%
%
%%%%%%%%%%%%%%%%%%%%%%%%%%%%%%
%
\bfi[ht!]
\bc\vspace{5mm}
\epsfig{file=int-v.eps,height=30mm}
\put(-87,42){$\gamma^*$}
\put(-0,-2){$\pi^0$}
\put(-0,80){$\phi$}
\put(-15,38){$F_{\phipiz}^{\rm Res}$}
\hspace{20mm}
\epsfig{file=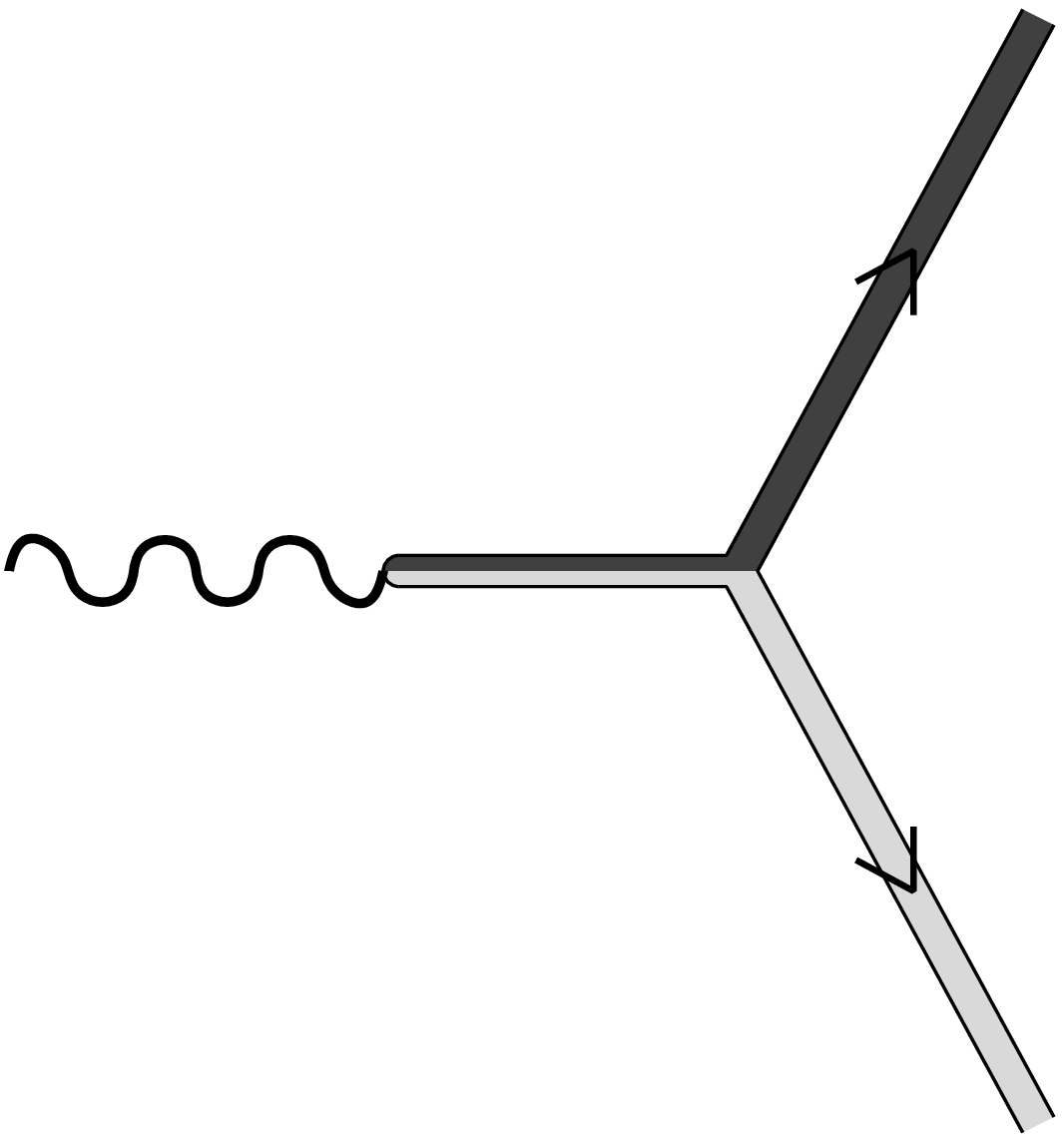,height=30mm}
\put(-120,42){$=\,\sum\limits_{j=1}^{N}$}
\put(-42,48){$V_j$}
\put(-68,26){
\bm{10mm}
$$
\frac{M_{V_j}^2}{F_{V_j}}
$$
\em
}
\put(-91,42){$\gamma^*$}
\put(-0,-2){$\pi^0$}
\put(-0,80){$\phi$}
\put(-15,38){$g^{V_j}_{\phipiz}$}
\vspace{0mm}
\ec
\caption{\label{fig:sum-v}
Schematic representation of the parameterization for the TFF in the resonance region.
}
\efi\vspace{5mm}\\
%
%%%%%%%%%%%%%%%%%%%%%%%%%%%%%%
%
      Such resonances are assumed to couple to the virtual photon 
      (see fig.~\ref{fig:sum-v}), hence the TFF is parametrized as a sum of propagators 
      weighted by the corresponding coupling constants:
      \be
      F^{\rm Res}_{\phipiz}=\sum_{j=1}^{N} \frac{M_j^2}{eF_{V_j}}\,
      \frac{g^{V_j}_{\phipiz}}{M_j^2-q^2-i\Gamma_j M_j}\;,
      \hspace{20mm} 4M_\pi^2\le q^2\le s^{\rm asy}\,,
      \label{eq:tff-res}
      \en
      where $M_j$ and $\Gamma_j$ are the mass and the width of the
      resonance $V_j$, $F_{V_j}$ and $g^{V_j}_{\phipiz}$ are the 
      couplings to the photon and to the hadronic final state. The number
      and species of intermediate resonances to be considered are established
      following two criteria: the quantum number conservation and the
      OZI rule. For this channel, being $I^G(J^{PC})=1^+(1^{--})$, we expect
      only $\rho$-like contributions even though they are OZI-suppressed. In particular
      we introduce three resonances: below threshold the $\rho^0(770)$, whose 
      parameters are completely fixed (see Sec.~\ref{sec:rho0}), above the physical
      threshold two additional $\rho$-recurrences, visible in the data 
      (fig.~\ref{fig:tff-data}): a broad structure around 1.6 GeV and a narrower 
      one at $\sim1.9$ GeV~\cite{babar}.
\item The asymptotic region (superscript ``asy''): 
      $q^2\ge s^{\rm asy}$, where we use the QcR power law
      behavior
\be
|F^{\rm asy}_{\phipiz}(q^2)|=|F^{\rm Res}_{\phipiz}(s^{\rm asy})|\left(
\frac{s^{\rm asy}}{q^2}
\right)^{n_h+\frac{n_\lambda+l_q-1}{2}=2},
\label{eq:tff-asy}
\en
$n_h=2$ is the number of external hadronic fields, $n_\lambda=1$
is the hadronic helicity, and $l_q=0$ is the quark-antiquark angular 
momentum in the $\pi^0$ (see Ref.~\cite{mia}).
\item The analytic region (superscript ``an''): 
      $q^2<4M_\pi^2$. In this region the TFF is reconstructed
      using the previous two expressions [eqs.~(\ref{eq:tff-res}) and~(\ref{eq:tff-asy})]
      in the dispersion relation integral~\cite{math,mia}:
      \be
      \!\!\!\!\!\!\!\!F^{\rm an}_{\phipiz}(q^2)\!=\!\exp\!\!\left[\!
	\frac{\sqrt{4M_\pi^2\!-\!q^2}}{\pi}\!\left(\!
      \int_{4M_\pi^2}^{s^{\rm asy}}\!\!\frac{\ln|F_{\phipiz}^{\rm Res}(s)|ds}
      {(s\!-\!q^2)\sqrt{s\!-\!4M_\pi^2}}\!+\!\!\!
      \int_{s^{\rm asy}}^\infty\!\!\frac{\ln|F_{\phipiz}^{\rm asy}(s)|ds}
      {(s\!-\!q^2)\sqrt{s\!-\!4M_\pi^2}}\!\right)\!\right]\!.
      \label{eq:def-an}
      \en
\ei
\section{Model independent description with only $\rho^0$ contribution}
\label{sec:rho0}
The $\rho^0$ contribution, due to the process $e^+e^-\to\gamma^*\to
\rho^0\to\phipiz$, represents an important OZI-violating intermediate
state. The presence of such a kind of contribution is inferred by
the large $\phi-\rho^0\pi^0$ coupling. The rate $\Gamma(\phi\to\rho^0\pi^0)$ 
for the single neutral channel $\rho^0\pi^0$ can be obtained from 
the total rate $\Gamma(\phi\to\rho\pi)$ correctly accounting for
the interference among the three isospin channels, for details see 
the Appendix~\ref{sec:app}. In particular,  we find that only the 18.6\% of
the BR$(\phi\to\rho\pi)=14.0\pm0.4$~\cite{pdg} is due to $\rho^0\pi^0$.
Using the general expressions:
\be
\Gamma(V\to V' P)&=&\frac{|g^V_{V'P}|^2}{12\pi}
\frac{
\left\{\left[M_V^2-(M_{V'}-M_{P})^2\right]
\left[M_V-(M_{V'}+M_{P})^2\right]
\right\}^{3/2}}{(2M_V)^3}\no\\
\Gamma(V\to e^+e^-)&=&\frac{\alpha M_{V}}{3 |F_{V}|^2}
\label{eq:g-coup}
\en
for the decay of a vector meson $V$ into another vector $V'$ and a pseudoscalar
$P$, and the electromagnetic coupling of a vector $V$ to $e^+e^-$ respectively,
we extract the couplings in terms of the corresponding rates and we get
($V=\phi$, $V'=\rho^0$, and $P=\pi^0$):
\be
|g^\phi_{\rho^0\pi^0}|=(0.865\pm 0.013)\,\rm GeV^{-1}\hspace{5mm}
\mbox{and}\hspace{5mm}
|F_{\rho^0}|=16.6\pm 0.2~.
\label{eq:rho-acco}
\en
The knowledge of these couplings completely determines the $\rho^0$ 
contribution in terms of the parameterization of eq.~(\ref{eq:tff-res}).
We are now ready to establish whether this contribution is enough to describe the
cross section measured by the \bbr\ Collaboration. Looking at fig.~\ref{fig:solorho0},
where the TFF and the corresponding cross section are shown as expected in case of
only $\rho^0$ contribution, we note that:
\bi
\item above the physical threshold $(M_{\pi^0}+M_\phi)^2$, the $\rho^0$ contribution
      alone is not enough to describe the data, indeed even though the measured 
      $\phipiz$ cross section is very small, less than 0.2 nb, this contribution
      gives solely few pb (see fig.~\ref{fig:solorho0}b);
\item below the theoretical threshold $q^2=4M_\pi^2$, and in particular at $q^2\!\!=\!0$,
      the prediction for the TFF: $F_{\phi\pi^0}(0)=(0.177\pm0.003)$~GeV$^{-1}$ is higher than 
      the experimental value extracted from the radiative decay rate 
      $\Gamma(\phi\to\pi^0\gamma)$ and reported in eq.~(\ref{eq:tff0}).
\ei 
%
%%%%%%%%%%%%%%%%%%%%%
%
\bfi[h!]
\bc
\epsfig{file=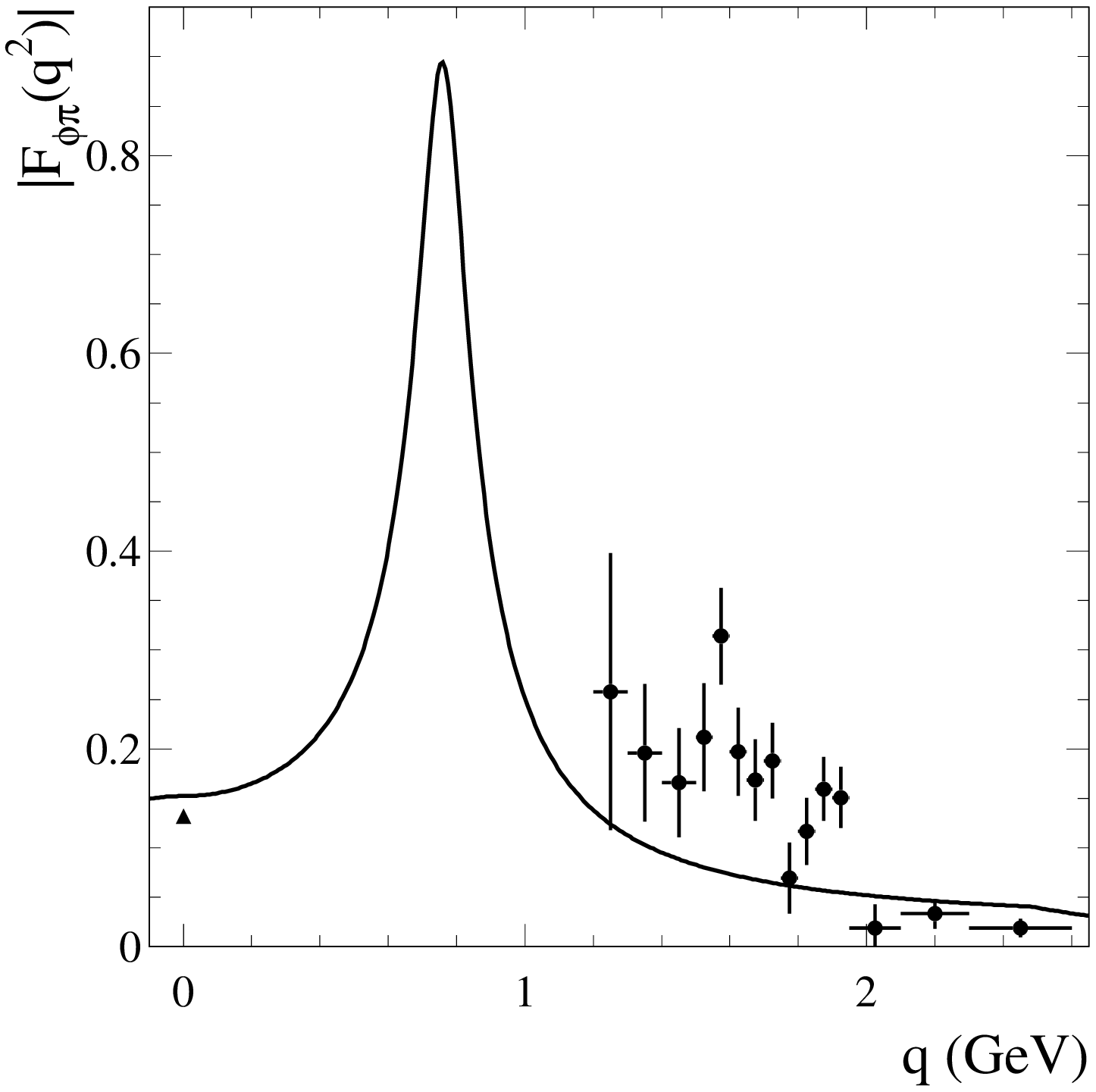,width=70mm}
\put(-165,180){\bf a}
\epsfig{file=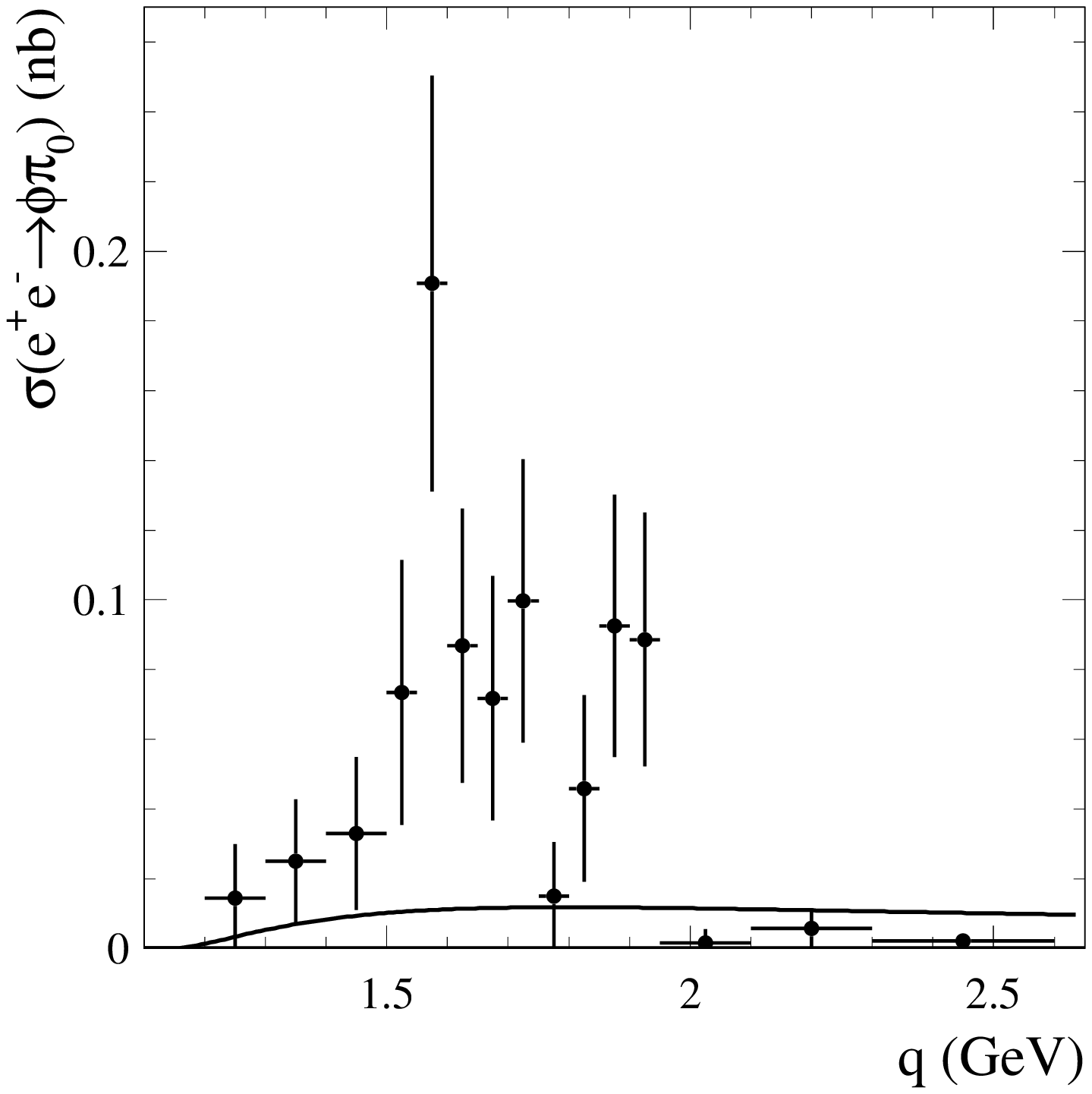,width=70mm}
\put(-165,180){\bf b}
\caption{\label{fig:solorho0}
Data (points) on the $\phipiz$ TFF (a) and the corresponding $e^+e^-\to\phipiz$ 
cross section (b) compared with the expectations in case of only  $\rho^0$ 
contribution (curves).}
\ec
\efi\vspace{-0mm}
%
%%%%%%%%%%%%%%%%
%
Both these observations point to the presence of additional
contributions. In light of the discrepancy between the predicted
TFF and data, we may deduce that these further resonances have to
lie around 1.6 and 1.9 GeV. Moreover, following the vector meson
dominance relation:
\be
g^\phi_{\pi^0\gamma}\equiv g^\gamma_{\phi\pi^0}\simeq \sum_{V} \frac{g^V_{\phi\pi^0}}{eF_{V}}
=\frac{g^{\rho^0}_{\phi\pi^0}}{eF_{\rho^0}} + \frac{g^{\rho'}_{\phi\pi^0}}{eF_{\rho'}}
+ \frac{g^{\rho''}_{\phi\pi^0}}{eF_{\rho''}}\;,
\label{eq:vmd-rel}
\en 
where two additional $\rho$-recurrences are included, to reconcile the
value of the TFF at $q^2=0$ with that of eq.~(\ref{eq:tff0}), $\rho'$
and $\rho''$ must give a global {\bf negative} contribution, i.e.:
\be
\frac{g^{\rho'}_{\phi\pi^0}}{eF_{\rho'}}
+ \frac{g^{\rho''}_{\phi\pi^0}}{eF_{\rho''}}\simeq -0.046\, .
\en
This means that at least the coupling of the $\rho'$, which looking at the data
of fig.~\ref{fig:solorho0} appears as the dominant contribution besides the $\rho^0$, 
must be negative.
\section{Global description including additional resonances}
\label{sec:global}
In this section we extend the analysis of the $\phipiz$ TFF including, in addition to 
the $\rho^0$, other resonant contributions as suggested by the arguments previously 
discussed. However, since the experimental information on the $\rho$-recurrences as well 
as their classification are quite uncertain and unstable, we consider masses, widths, 
and couplings of the additional contributions as free parameters to be fit to the
data and to the theoretical constraints. Therefore, contrary to the previous case
where, with the only $\rho^0$ contribution, we had a fully frozen parametrization, 
this global description is now model-dependent.
\subsection{$\rho(1450)$, $\rho(1700)$ and $C(1480)$}
\label{sec:rho1450}
Since 1988 the Particle Data Group~\cite{pdg} decided to replace the $\rho(1600)$,
that were the only excited $\rho$ below 2 GeV, with two states: $\rho(1450)$ 
and $\rho(1700)$. This is a consequence of the clear theoretical and experimental
evidence that the 1.6 GeV energy region contained more than one $\rho$-recurrence.  
Furthermore, also the so-called $C(1480)$ observed in the charge exchange reaction 
$\pi^- p\to\phi\pi^0 n$ as a clear peak in the $\phi\pi^0$ invariant mass 
distribution~\cite{Bityukov:1986yd}, has to be included in the list of all
possible contributions. The observation of this structure generated
a great deal of theoretical and experimental 
works~\cite{c1480tutto}. In particular, there has been a theoretical effort
to find the best observable to identify the nature of this resonance:
$q\ov{q}$-$\rho$-like meson or exotic with a hidden strangeness content. 
The electronic width $\Gamma(C(1480)\to e^+e^-)$ and the branching 
BR$(C(1480)\to\phi\pi^0)$ seemed good candidates for this purpose. The
leptonic width for a tetraquark state should be suppressed with respect to that
for a $q\ov{q}$ by a factor $\ll 1$, representing the price to be paid in creating
an extra quark pair from the vacuum. Instead, the coupling of an exotic $C(1480)$
to the $\phi\pi^0$ final state should be favored compared with that of
a standard meson, which is OZI-suppressed. Unfortunately the cross section
$\sigma(e^+e^-\to C(1480)\to\phi\pi^0)$ gives information only on the product
of the two couplings, that, without data on other independent channels,
cannot be disentangled. It follows that the cross section alone does not
allow to distinguish between two possible resonant contributions: the exotic $C(1480)$ 
and the standard $\rho(1450)$, if both exist. Moreover, there are authors
claiming that the $C(1480)$ is actually the $\rho(1450)$~\cite{Achasov:1987ku}.
\\
The overpopulation of this energy region makes the description of TFF's in terms
of resonances difficult. The lack of information on partial widths for these
$\rho$-recurrences and the impossibility of distinguishing their contributions in
certain decay channels puzzle the extraction of the couplings, which is essential
for this description. In the following we will refer to the first excited $\rho$ 
of the $\phi\pi^0$ TFF simply as the $\rho'$, avoiding any a priori identification
with  $\rho(1450)$, $C(1480)$, or $\rho(1700)$ and gaining precious information from the one 
or the other according to the case.
\\
For instance, the first important hint on the coupling with the actual final state 
$\phipiz$ can be obtained for the $\rho(1450)$. Even though there is no 
direct information on the $g^{\rho'}_{\phipiz}$, as in the $\rho^0$ case 
(Sec.~\ref{sec:rho0}), the known upper limit~\cite{pdg,upper}:
\be
\Gamma(\rho'\to e^+e^-){\rm BR}(\rho'\to\phi\pi)\equiv
\frac{\Gamma(\rho'\to\phi\pi)\Gamma(\rho'\to e^+e^-)}{\Gamma_{\rho'}}<
70\;\rm eV 
\label{eq:upper}
\en
helps in understanding whether $\rho(1450)$, here called $\rho'$, can contribute 
to the annihilation $e^+e^-\to\phipiz$.\\
Multiplying the two expressions of eq.~(\ref{eq:g-coup}), with $V=\rho'$, 
$V'=\phi$, and $P=\pi^0$ we obtain
\be
\!\!\!\!\frac{\Gamma(\rho'\!\!\to\!\!\phi\pi)\Gamma(\rho'\!\!\!\to\! e^+\!e^-)}{\Gamma_{\rho'}}
=
\frac{|g^{\rho'}_{\phi \pi^0}|^2}{12\pi\Gamma_{\rho'}}
\frac{\left\{\!\left[\!M_{\rho'}^2\!-\!(\!M_{\phi}\!-\!M_{\pi^0}\!)^2\!\right]\!\!
\left[\!M_{\rho'}^2\!-\!(\!M_{\phi}\!+\!M_{\pi^0}\!)^2\!\right]
\!\right\}^{\frac{3}{2}}}{(2M_{\rho'})^3}
\frac{\alpha M_{\rho'}}{3|F_{\rho'}|^2}\,,
\en
which, using the values for $M_{\rho'}$ and $\Gamma_{\rho'}$ given in Ref.~\cite{pdg},
provides the upper limit:
\be
\Bigg|\frac{M_{\rho'}^2g^{\rho'}_{\phi \pi^0}}{F_{\rho'}}\Bigg|< 0.19\;\rm GeV.
\en
This value leaves room to an important contribution from the $\rho'$,
 it is quite high with respect to the corresponding coupling for the
$\rho^0$:  $|M_\rho^2 g^{\rho}_{\phi \pi^0}/F_{\rho}|\simeq0.03\;\rm GeV^{-1}$.
In particular, the peak values are [see eq.~(\ref{eq:tff-res})]:
\be
\left|\frac{M_{\rho}g^{\rho}_{\phi\pi^0}}{e F_{\rho}\Gamma_{\rho}}\right|\simeq 0.9\;{\rm GeV}^{-1}
\hspace{10mm}\mbox{and}\hspace{10mm}
\left|\frac{M_{\rho'}g^{\rho'}_{\phi\pi^0}}{e F_{\rho'}\Gamma_{\rho'}}\right|<
1.1\;{\rm GeV}^{-1} .
\en
%
%%%%%%%%
%
\subsection{Once more the $\rho(1900)$}
\label{sec:rho1900}
The cross section data, as well as the data for the TFF reported 
in fig.~\ref{fig:tff-data}, show an accumulation of events around 1.9 GeV.
Following the line of thought of Ref.~\cite{babar}, we identify this
structure as the second excited $\rho$, called $\rho''$ or $\rho(1900)$.
In the same reference the statistical significance for such a resonance
is found to be not weak being $2\times 10^{-3}$, which means 
there is a 0.2\% chance that the result was accidental.
\\ 
It appears as an already known structure, previously observed in 
six-$\pi$~\cite{six} and four-$\pi$~\cite{four} final states. 
As suggested in Ref.~\cite{exotic} it could be the manifestation of a 
cryptoexotic $J^{PC}=1^{--}$ tetraquark state, whose coupling with a
 multi-particle final state should be favored by a milder suppression 
factor due to the creation of a lower number of quark pairs from the 
vacuum.
\\
If the bump in the \bbr\ $\phipiz$ cross section is the already known
$\rho(1900)$, this should be the first observation of $\rho(1900)$ decaying
in this channel and 
also its first manifestation as a peak instead of a dip.
%
%
%%%%%%%%
%
\subsection{$\chi^2$ definition}
\label{sec:fit}
We are now ready to define the $\chi^2$. We work directly with the TFF instead of the
cross section, this allows us to take advantage from important regularities, as analyticity,
and constraints from different sets of data. The $\chi^2$ is defined
as the sum of two contributions. The first one is from the data on both the 
annihilation cross section $\sigma(e^+e^-\to \phi\pi^0)$ and the radiative decay 
rate $\Gamma(\phi\to \pi^0\gamma)$, and the second comes from the theory, i.e. 
the analyticity requirement, which is imposed by means of the DR's. \\
In more detail we have:
\be
\chi^2=\chi_{\rm exp}^2+\tau\cdot\chi_{\rm th}^2\; ,
\label{eq:chitot}
\en
where $\tau$ is a weighting factor for the theoretical part,
that will be described in the following.
The experimental contribution is:
\be
\chi_{\rm exp}^2=\sum_{j=1}^N\left(\frac{|F_{\phi\pi^0}(s^{\rm exp}_j)|-
F^{\rm exp}_j}{\delta F^{\rm exp}_j}\right)^2+
\left(\frac{|F_{\phi\pi^0}(0)|-F^{\rm exp}_0}{\delta F^{\rm exp}_0}\right)^2.
\en
The first term concerns the annihilation data $\{s^{\rm exp}_j,F^{\rm exp}_j\}$,
with: $j=1,2,\ldots,N$ and $s^{\rm exp}_j\ge (M_\phi+M_\pi)^2$, while the
second is for the radiative decay rate.
The theoretical component $\chi_{\rm th}^2$ forces the analyticity
of the TFF through the DR for the logarithm~\cite{drlog,mia}
\be
\ln F(q^2)=\frac{\sqrt{4M_\pi^2-q^2}}{\pi}\int_{4M_\pi^2}^\infty
\frac{\ln|F(s)|ds}{(s-q^2)\sqrt{s-4M_\pi^2}}~,
\label{eq:dr}
\en 
that gives the real value of the TFF below the theoretical threshold 
($q^2<4M_\pi^2$) in terms of an integral of the logarithm of its modulus
over the cut $(4M_\pi^2,\infty)$. The relation of eq.~(\ref{eq:dr})
guarantees the continuity of the parameterization for the TFF 
(zero-th derivative) across the threshold 
$4M_\pi^2$. Instead, the continuity of the first derivative has to be imposed
to connect the last two definitions of eq.~(\ref{eq:3def}), in the resonance
and asymptotic regions, to the first one in the analytic region.
In formulae we get:
\be
\chi^2_{\rm th}=\left(\left.\frac{dF_{\phi\pi^0}}{ds}\right|_{s=s_{\rm th}}
-\int_{s_{\rm th}}^\infty K(q^2\to s_{\rm th}^+,s)\ln|F_{\phi\pi^0}(s)|ds\right)^2,
\en
where $s_{\rm th}=4M_\pi^2$ and $K(q^2,s)$ is the derivative of the kernel of eq.~(\ref{eq:dr}):
\be
K(q^2,s)=\frac{2s_{\rm th}-s-q^2}{2\pi\sqrt{s_{\rm th}-q^2}\sqrt{s-s_{\rm th}}(s-q^2)^2}~.
\en
Since this condition has to be exactly verified, the value of the parameter 
$\tau$, which weights the $\chi^2_{\rm th}$ in eq.~(\ref{eq:chitot}), must be chosen
large enough to force the vanishing of the $\chi^2_{\rm th}$ itself.
%
%%%%%%%%%%
%
\subsection{Fit results}
\label{sec:result}
\begin{figure}[h]
\epsfig{file=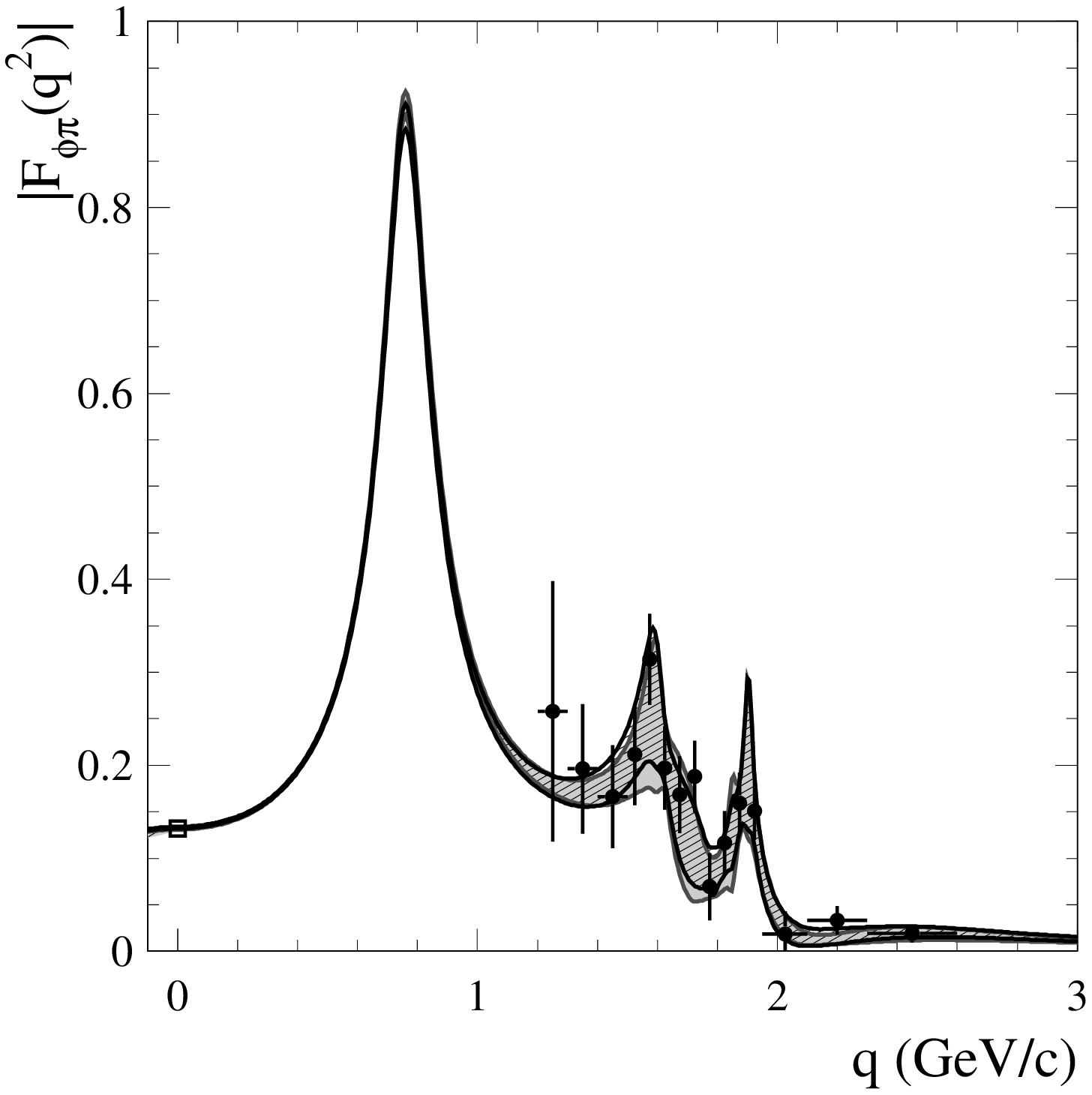,width=73mm}
\put(-170,185){\bf a}\hfill
\epsfig{file=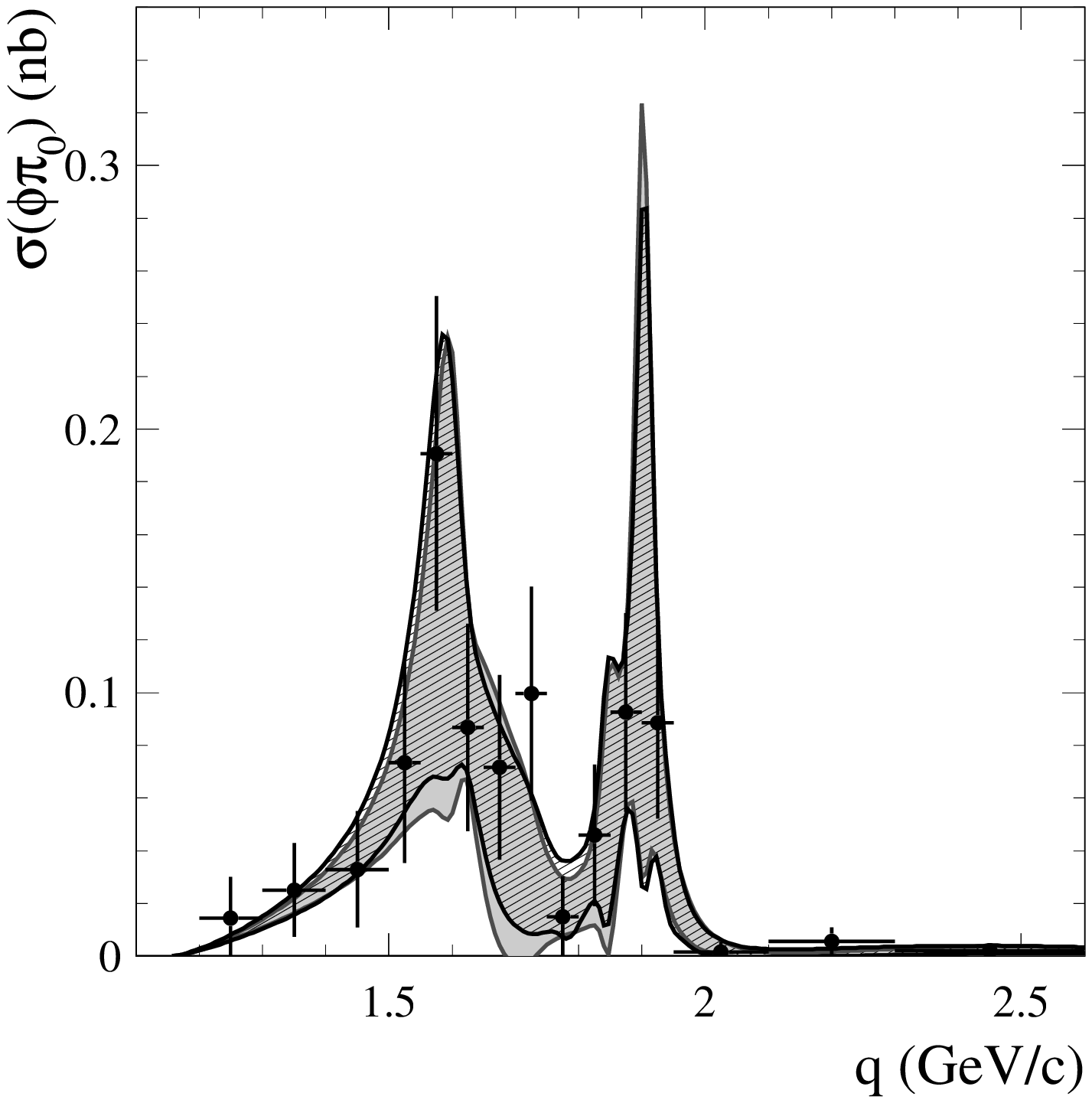,width=73mm}
\put(-170,185){\bf b}
\caption{\label{fig:fits}
The fit of the $\phipiz$ TFF (a) and of the corresponding cross
section (b). The error bands have been obtained with a Monte Carlo 
technique (see the text). The solid gray band corresponds to the 
case where we use real couplings, while the hatched one is for
complex couplings.
}
\end{figure}

Before giving the result of the fit, we summarize the key points of the 
parameterization used for the $\phipiz$ TFF. We adopted the threefold
definition of eq.~(\ref{eq:3def}) given in Ref.~\cite{mia}.
The time-like region, where we parametrize the TFF, has been divided in
three intervals. The asymptotic region, from a certain energy $s^{\rm asy}$
(free parameter in the fit) up to infinity, where we used the QcR prescription which gives the
power-law behavior: $|F_{\phipiz}(q^2)|\propto (q^2)^{-2}$ as $q^2\to\infty$.
The resonance region $(4M_\pi^2,s^{\rm asy})$, where the TFF is described in 
terms of vector meson propagators. More in detail in that region we consider the three resonances
$\rho^0$, $\rho'$, and $\rho''$, and the TFF is:
\be
F_{\phi\pi^0}^{\rm Res}(s)&=&\frac{M_{\rho}}{eF_{\rho}}\frac{{g^{\phi}_{\rho\pi^0}}}{\Gamma_\rho}
\frac{M_{\rho}\Gamma_\rho}{M_{\rho}^2-s-i\Gamma_\rho M_\rho}+
\frac{{M_{\rho'}}}{eF_{\rho'}}\frac{
g^{\phi}_{\rho'\pi^0}}
{{\Gamma_{\rho'}}}
\frac{{M_{\rho'}}{\Gamma_{\rho'}}}{{M_{\rho'}}^2-s-i
{\Gamma_{\rho'}} {M_{\rho'}}}e^{i\delta}+\no\\
\={}+\frac{{M_{\rho''}}}{eF_{\rho''}}\frac{
g^{\phi}_{\rho''\pi^0}}
{{\Gamma_{\rho''}}}
\frac{{M_{\rho''}}{\Gamma_{\rho''}}}{{M_{\rho''}}^2-s-i
{\Gamma_{\rho''}} {M_{\rho''}}}e^{i\delta'} +A_{\rm n.r.}\;,
\label{eq:new-TFF3}
\en
where we include also additional relative phases $\delta'$ and $\delta''$
to account for possible rescattering effects, and a non-resonant 
constant amplitude $A_{\rm n.r.}$. Finally, the analytic region
$q^2<4M_\pi^2$, where the real TFF is obtained using the previous two definitions
in the integral of the DR for the logarithm of eq.~(\ref{eq:dr}).
\\
We will consider two fits: with real couplings, i.e.: the phases are 
set to zero and the $g^V_{\phipiz}$'s are allowed to take negative and 
positive values (gray bands in fig.~\ref{fig:fits}), and with complex couplings, i.e. with free
relative phases (hatched bands in fig.~\ref{fig:fits}).
%
%%%%%%%%%%%%%%%%%%%%%%%%%%%
%
\\
%
%%%%%%%%%%%%%%%%%%%%%%%%%%%
%
The fit is shown in fig.~\ref{fig:fits} and the corresponding
best values for masses, widths, phases, and total couplings are reported in 
table~\ref{tab:best}. The error bands have been obtained
with the following procedure: we generate many samples of data by Gaussian fluctuation
of the original ones, we fit all these samples obtaining  sets of parameters and
 curves for the TFF, we treat these parameters and curves with the usual
statistical procedures. We take the mean and the standard deviation as the best 
value and the corresponding error for the parameters and, as the central value 
and the half-width of the band at each $q\equiv \sqrt{q^2}$ for the TFF and
cross section curves (see fig.~\ref{fig:fits}).\\
The parameters achieved with this technique are in agreement with those obtained 
in Ref.~\cite{babar} with a different fit procedure, that does not exploit information 
coming from the radiative decay rate and the QcR for the asymptotic behavior. 
This is the reason why, in case of real
couplings, we get the same sign for $\rho'$ and $\rho''$ while they are opposite in 
Ref.~\cite{babar}, where there is a lower sensitivity to the complex structure 
of the fit function.\\
The additional flat non-resonant contribution turns out to be only a small
fraction (about 2.5\% of the $\rho^0$ peak value) of the TFF in the resonance region.
%%%%%%%%%%%%%%%%%%%%%%%%%%%%
%
\begin{table}[h!]
\bc
\renewcommand{\arraystretch}{1.5}
\begin{tabular}{|c|c|c|c|c|c|}
\hline
\rm Cont. & $\Big|\frac{M_{V}^2}{F_{V}}g^{\phi}_{V\pi^0} \Big|$
(MeV) & $\Gamma^V_{e^+e^-}$BR$^{V}_{\phipiz}$ (eV) &\rm $M$ (MeV) & \rm $\Gamma$ (MeV) & \rm Phase (rad) \\
\hline
\multirow{2}{*}{$\rho'$} & $(-)15\pm 6$ & $2.8\pm 0.9$ & $1640 \pm 46$ & $143\pm 85$ & $\pi(0)$ \\
\cline{2-5}
  & $16\pm 7$ & $3.1\pm1.4$ & $1617 \pm 58$ & $150\pm 99$ & $3.0\pm 0.5$ \\
\hline
\multirow{2}{*}{$\rho''$} & $(-)6.8\pm2.9$ & $2.3\pm 0.8$ & $1895\pm22$ & $62\pm46$ & $\pi(0)$ \\
\cline{2-5}
   & $7.3\pm3.5$ & $2.6\pm 1.0$ & $1895\pm29$ & $67\pm55$ & $3.5\pm 0.8$ \\
\hline
\hline
\multirow{2}{*}{$A_{\rm n.r.}$ } &  \multicolumn{3}{c|}{$-0.024\pm 0.006$ GeV$^{-1}$} & 
\multicolumn{2}{c|}{\multirow{2}{*}{$\simeq 0.025\cdot |F_{\phipiz}(M_\rho^2)|$}}\\
\cline{2-4}
  &  \multicolumn{3}{c|}{$-0.023\pm 0.006$ GeV$^{-1}$} & \multicolumn{2}{c|}{}\\
\hline
\end{tabular}
\caption{Best fit parameters. 
For each parameter we give two values corresponding to the cases where
we use real couplings, upper rows, and complex couplings (additional relative
phases), lower rows. The value of the constant amplitude is also given in terms 
of the TFF at the $\rho^0$ peak, and, in both cases it represents about the 
$2.5\%$ of this peak value.}
\label{tab:best}
\ec
\vspace{-5mm}
\end{table}
%
%
%%%%%%%%%%%%%%%%%%%%%%%%%%%
%
\subsection{$g^{\rho'}_{\phipiz}$ coupling}
\label{sec:rhoprime}
Since $\rho^0$, $\rho'$, and $\rho''$ are intermediate resonances produced through 
an $e^+e^-$ annihilation process, the cross section can give information only on the 
product between the electromagnetic and the mesonic couplings, e.g.: the modulus of
$(M_V^2/F_V)g^V_{\phipiz}$ for a generic vector meson $V$ (see table.~\ref{tab:best}
for $\rho'$ and $\rho''$).
To know the pure mesonic coupling $g^{V}_{\phipiz}$, which describes the transition 
$\phi\to\pi^0$ with emission of a virtual vector meson $V$, we need to know the
coupling $F_{V}$, which instead gives the amplitude of the decay $V\to e^+e^-$. \\
The situation appears quite interesting in the case of the $\rho'$ resonance.
This structure is mainly observed decaying in four-pion final states,
hence, assuming such final states as the dominant decay channels we can give
an estimate, or at least an upper limit, for $F_{\rho'}$. Using the usual
Breit-Wigner formula for a generic process: $e^+e^-\ra V\ra \rm$ [final state],
which proceeds through an intermediate vector resonance $V$, we have
at the vector meson peak:
\be
\sigma(\mbox{$V$ peak})=12\pi\frac{\Gamma(V\ra e^+e^-)\cdot 
{\rm BR}(V\ra \mbox{[final state]})}{M_V^2\Gamma_V}\,,
\label{eq:bw}
\en
where $M_V$ and $\Gamma_V$ are the mass and total width of the resonance.
Being the branching BR$(V\ra \mbox{[final state]})$ lower or equal to one
(it is $\simeq 1$ when the final state is the dominant decay channel),
from eq.~(\ref{eq:bw}) we get the upper limit
\be
\Gamma(V\ra e^+e^-)\le\frac{\sigma(\mbox{$V$ peak})M_V^2\Gamma_V}{12\pi}~.
\label{eq:upper1}
\en
It remains to relate the decay rate $\Gamma(V\ra e^+e^-)$ to the coupling
$F_V$, by means of eq.~(\ref{eq:g-coup}) we obtain:
\be
|F_V|=\sqrt{
\frac{\alpha M_V}{3\Gamma(V\ra e^+e^-)}
}\ge\sqrt{
\frac{4\pi\alpha}{\sigma(\mbox{$V$ peak})M_V\Gamma_V}
}=\frac{e}{
\sqrt{\sigma(\mbox{$V$ peak})M_V\Gamma_V}}~.
\label{eq:sig-peak}
\en
%%%%%%%
%%%Sec.~\ref{sec:rho1450}
%
If we identify the $\rho'$ with the $\rho(1700)$ of Ref.~\cite{pdg},
we may take advantage from the fact that the four-pion final state
is the dominant decay channel. Under this hypothesis we estimate the 
cross section at the $\rho'$ peak.
There are two channels: $\rho'\to 2(\pi^+\pi^-)$ 
and $\rho'\to 2\pi^0 \pi^+\pi^-$ and the peak values of the cross sections
are~\cite{four}:
\be
\left.\begin{array}{l}
\sigma_{2(\pi^+\pi^-)}(\rho'\;\rm peak)= 29.2\pm 0.7\,{\rm nb}\\
\\
\sigma_{\pi^+\pi^- 2\pi^0}(\rho'\;\rm peak)= 18.2\pm 0.7\,{\rm nb}\\
\end{array}\right\}
\hspace{5mm}\Longrightarrow\hspace{5mm} \sigma_{4\pi}(\rho'\;\rm peak)= 47.4\pm 1.0\;{\rm nb}~.
\en
From the formula of eq.~(\ref{eq:sig-peak}) and using the value for 
$|M_{\rho'}^2g^{\rho'}_{\phipiz}/F_{\rho'}|$  reported in table~\ref{tab:best}
we get:
\be
\left|F_{\rho'}\right|\ge\left\{ 
\begin{array}{l}
65\pm 20\\
\\
64\pm 18\\
\end{array}
\right.\hspace{10mm}
\mbox{and}\hspace{10mm}
\left|g^\phi_{\rho'\pi^0}\right|\le
\left\{ 
\begin{array}{l}
0.32\pm 0.07\; {\rm GeV}^{-1}\\
\\
0.36\pm 0.07\; {\rm GeV}^{-1}\\
\end{array}
\right.
,
\label{eq:accoppia}
\en
where, as in the following, the upper and lower values refer to the cases with real 
and complex couplings respectively. The electromagnetic branching ratios are:
\be
{\rm BR}(\rho'\ra e^+e^-)\le\frac{\sigma_{4\pi}(\mbox{$\rho'$ peak})M_{\rho'}^2}{12\pi}
=
\left\{ 
\begin{array}{l}
(8.7\pm 0.5)\times 10^{-6}\\
\\
(8.5\pm 0.6)\times 10^{-6}\\
\end{array}
\right.
.
\en
Finally, by using the value for $\Gamma(\rho'\to e^+e^-)$BR$(\rho'\to\phipiz)$, 
we can extract upper limits for the rates of the decay $\rho'\ra\phi\pi^0$ as
\be
\Gamma(\rho'\ra\phi\pi^0)\le
\left\{ 
\begin{array}{l}
330\pm 115\;{\rm keV}\\
\\
360\pm 150\;{\rm keV}\\
\end{array}
\right.
.
\en
These values are lower than the OZI-violating width of the $\phi$ meson in its
decay in $\rho\pi+3\pi$, which is:
\be
\Gamma(\phi\to\rho\pi+3\pi)=(0.153\pm 0.004)\cdot (4260\pm 50)\;{\rm keV}=650\pm 20\;{\rm keV}.
\en
%
%
%%%%%%%%%%%%%%%%%%%%%%%%%%%
%
\section{Conclusions}
\label{sec:conclu}
The $\phi\pi^0$ TFF has been studied by means of a dispersive
procedure which collects all the experimental information under the 
aegis of the analyticity. The main low-energy contribution to this TFF 
is provided by the $\rho^0$ meson whose coupling to $\phi\pi^0$
is well known. If we assume only this contribution we get a 
model-independent description without free parameters.
However, as shown in fig.~\ref{fig:solorho0}, this single-resonance TFF 
is unable to describe neither the 
data from the cross section $\sigma(e^+e^-\to\phi\pi^0)$, nor the point 
at $q^2=0$ from the radiative decay rate $\Gamma(\phi\to\pi^0\gamma)$.
\\ 
Therefore we considered, in addition to the $\rho^0$, two further resonances defined 
as $\rho$-recurrences, that, having free parameters, make the description model-dependent.
The first excited $\rho$-recurrence of our analysis, 
the $\rho'$, has mass and width compatible ($\Delta M$, $\Delta\Gamma<$ two 
standard deviations) with those of the $\rho(1700)$ of Ref.~\cite{pdg}: 
$$
M_{\rho(1700)}=1720\pm 20\;{\rm MeV} \hspace{10mm} \Gamma_{\rho(1700)}=250\pm100\;{\rm MeV}
$$
In addition in the first case, assuming the four-pion final state as the dominant
decay channel, we have estimated the electromagnetic coupling $F_{\rho'}$
and the $g^{\rho'}_{\phipiz}$ [eq.~(\ref{eq:accoppia})]. The obtained
values are compatible with the expectation for an excited $\rho$-like state, given
those of the ground state $\rho^0$ reported in eq.~(\ref{eq:rho-acco}).
\\
The second excited state, the $\rho(1900)$, is perfectly compatible with
the ``dip'' observed for the first time by the DM2 Collaboration in the
six-charged pion final state and subsequently confirmed by other experiments
in the same final state~\cite{six} and also in the four-pion channel~\cite{four}.
In all these cases it appeared as a dip, this could be the first manifestation
as a peak. There are no clear interpretations for this structure, but the
most interesting, that could explain the coupling with the OZI-forbidden
$\phipiz$ final state and the multi-particle decay channels, is a vector meson 
cryptoexotic state~\cite{exo}, i.e.: hybrid $[q\ov{q}g]$ or
tetraquark $[qq\ov{qq}]$.
\\
Finally, we may conclude that, with the present statistics, there is no evidence 
of the $C(1480)$, i.e. the $\phipiz$ TFF is well described by the three main 
contributions: $\rho^0$, $\rho(1700)$, and $\rho(1900)$. In addition, 
the absence of further structures is also theoretically enforced by the
dispersive procedure used to describe the data. 
%
%%%%%%%%%%
%
\acknowledgments{%
I would like to express my gratitude to professor Antonino Zichichi and to the Enrico Fermi Center, for their
strong support, and to Rinaldo Baldini, Lia Pancheri, Yogi Srivastava
and Adriano Zallo for the precious and instructive discussions 
which helped me in developing this work.%
}
%
%%%%%%
\newpage
\appendix
\section{Rate $\Gamma(\phi\to\pi^0\rho^0)$}
\label{sec:app}
The $\phi$ meson decays into $\pi^+\pi^-\pi^0$ with a branching ratio of 
$\simeq 15.5\%$, however this decay is dominated by the $\rho\pi$ itermediate
states. The three channels: $\rho^0\pi^0$, $\rho^+\pi^-$ and 
$\rho^-\pi^+$ participate with equal amplitude.\\
This process has been recently studied by the Kloe Collaboration~\cite{cesare} 
and we will use their fit
procedure in order to reconstruct the Dalitz plot and then to extract
the rate for the single channel $\phi\to\rho^0\pi^0$.\\
The observed decay is $\phi(p_\phi)\to \pi^0(p_0)\pi^+(p_+)\pi^-(p_-)$, where the
4-momenta in parentheses are in the center of mass of the $\phi$.
The Dalitz plot density distribution is:
\be
D(x,y)\propto |\vec{p}_+\times \vec{p}_-|^2\cdot|A_{\rho\pi}+A_{\rm dir}+A_{\omega\pi}|^2\,,
\label{eq:dalitz}
\en
where $x$ and $y$ variables are:
\be
x=E_+ -E_-\,,\hspace{30mm}y=M_\phi-M_{\pi^0}-E_+ -E_-\,,
\en
and the three amplitudes $A_\alpha$ ($\alpha=\rho\pi,\,\rm dir,\,\omega\pi$) are:
\be
A_{\phi\pi}\!&=&\!\sum_{k}^{0,+,-}a_k
\frac{M_k^2}{q_k^2-M_k^2+iq_k\gamma_k(q_k^2)}\,,
\hspace{20mm}
\gamma_k(q_k^2)=\Gamma_k
\left(\frac{q_k^2-s_{0k}^2}{M_k^2-s_{0k}^2}\right)^\frac{3}{2}
\frac{M_k^2}{q_k^2}\,,\no\\
A_{\omega\pi}\!&=&\!
a_\omega e^{i\phi_\omega}\frac{M_\omega^2}{q_0^2-M_\omega^2+iq_0\Gamma_\omega}\,,\\
&&\no\\
A_{\rm dir}\!&=&\!a_{\rm dir}e^{i\phi_{\rm dir}}\no\,.
\en
The quantity $q_{k}^2$ ($k=0,+,-$) is the invariant mass of the pion pair with 
electric charge $k$:
\be
q_0^2\!&=&\!(p_+ +p_-)^2=(p_\phi -p_0)^2=(M_\phi-M_{\pi^0})^2-2M_\phi y\no\\
q_+^2\!&=&\!(p_0 +p_+)^2=(p_\phi -p_-)^2=M_\pi^2+M_\phi(M_{\pi^0}+x+y)
\\
q_-^2\!&=&\!(p_0 +p_-)^2=(p_\phi -p_+)^2=M_\pi^2+M_\phi(M_{\pi^0}-x+y)\,,
\no
\en
then $M_k$ and $\Gamma_k$ are masses and widths of the meson $\rho^k$,
and the values $s_{0k}$ represent the corresponding thresholds:
\be
s_{00}=4M_\pi^2\,, \hspace{3cm}
s_{0+}=s_{0-}=(M_\pi+M_{\pi^0})^2\,.
\en
We consider now the fractions:
\be
R^\alpha_\beta=\frac{\int dx dy |A_\alpha(x,y)|^2}{\int dx dy|A_{\beta}(x,y)|^2}\,,
\en
where $\alpha$ and $\beta$ are either single channels or sums of 
channels. By taking the parameters directly from Ref.~\cite{cesare} [Fit(c)]
we get:
\be
R_{\rho\pi}^{\rho^0\pi^0}=
\frac{\int dx dy |A_{\rho^0\pi^0}(x,y)|^2}{\int dx dy|A_{\rho\pi}(x,y)|^2}
=0.187\,.
\label{eq:R0}
\en
This number is quite far from the natural expectation: 1/3, this is due to
large interference effects among the three $\rho$'s in the Dalitz plot 
density [see eq.~(\ref{eq:dalitz})].
\section*{}
%
%%%%%%%%%%%%%%%%%%%%%%%%%%%%%%%%%%%%%%%%%%%%%%%%%%%%%%%%%%%%%%%%%%%%%%%%

%
%
%%%%%%%%%%%%%%%%%%%%%%%%%%%%%%%%%%%%%%%%%%%%%%%%%%%%%%%%%%%%%%%%%%%%%%%
\end{document}